\documentclass[a4paper]{iopart}

\usepackage{amsfonts,amsgen,amsbsy,amssymb,mathrsfs}
\usepackage{hyperref}

\def\>{\rangle}\def\<{\langle}
 \def\mE{\mathcal{E}}
 \def\fid{\mathsf{F}} \def\N#1{\left|\!\left|{#1}\right|\!\right|}
\def\supp{\textrm{Supp}} \def\instr{\mathcal{M}}

\def\out{\boldsymbol{\mathcal{X}}} \def\mR{\mathcal{R}}
\def\X{\boldsymbol{X}} \def\id{\textrm{id}}

\def\s#1{\boldsymbol{\mathcal{#1}}} \def\mS{\mathfrak{s}}
\def\cfid{\overline{F_{av}}}
\def\cFe{\overline{F_e}}
\def\Iacc{I_\mathrm{acc}}
\def\x{\boldsymbol{x}}
\DeclareRobustCommand\openone{\leavevmode\hbox{\small1\normalsize\kern-.33em1}}

\newtheorem{lemma}{Lemma}
\newtheorem{theo}{Theorem}

\newtheorem{prop}{Proposition}

\begin{document}

\title[Towards a unified approach to information-disturbance
tradeoffs]{Towards a unified approach to information-disturbance
  tradeoffs in quantum measurements}

\author{Francesco Buscemi} 
\address{ERATO-SORST Quantum Computation and Information Project,
  Japan Science and Technology Agency}
\ead{buscemi@qci.jst.go.jp}

\author{Micha\l{} Horodecki} \address{Institute of Theoretical
  Physics and Astrophysics, University of Gda\'nsk, Poland}

\date{\today}

\begin{abstract}
  We show that the global balance of information dynamics for general
  quantum measurements given in [F.~Buscemi, M.~Hayashi, and
  M.~Horodecki, Phys.~Rev.~Lett.~{\bf 100}, 210504 (2008)] makes it
  possible to unify various and generally inequivalent approaches
  adopted in order to derive information-disturbance tradeoffs in
  quantum theory. We focus in particular on those tradeoffs,
  constituting the vast majority of the literature on the subject,
  where disturbance is defined either in terms of average output
  fidelity or of entanglement fidelity.
\end{abstract}

\maketitle

\section{Introduction}

The general idea for which information extraction always causes
disturbance is nowadays widely accepted as one of the most fundamental
and distinctive principles of quantum theory as opposed to classical
theory. However, without precisely defining what we mean with the
vague words ``information'' and ``disturbance'', such a statement is
nothing but an empty sentence. In order to obtain some mathematically
sound results, one first needs to provide a natural and meaningful way
to measure both information and disturbance. Given these definitions,
an information-disturbance tradeoff relation is then a lower bound on
(some monotonic function of) disturbance given in terms of (some other
monotonic function of) information gain, in such a way that the
conclusion ``disturbance is null only if information gain is null''
can be drawn from it. With this at hand, a second order of problems is
to identify \emph{least-disturbing measurements}, namely, those
measurements causing the minimum disturbance compatible with a given
amount of information extracted. For these kind of ``optimal''
measurements then one would like to see that also the converse is
true, that is, if information gain is null, then also minimum
disturbance is null, thus establishing an equivalence relation between
information and minimum disturbance. This is however beyond the scope
of the present paper: our main concern will be to unify different and
usually inequivalent ways to quantify information and disturbance,
hoping that such a clarification could simplify the derivation of
optimal measurements, which in general is an awkward, yet important,
task.



The setting of the problem is usually as follows: a letter $x$, drawn
from an input alphabet $\X$ according to the probability distribution
$p(x)$, is encoded into some quantum state $\varrho_x$. Subsequently,
a quantum measurement $\instr$ (for the moment, let us think of the
measurement as a kind of black box performing some fixed operation) is
performed on the given state $\varrho_x$, whose label $x$ is unknown,
producing a measurement readout letter $m$, belonging to the set
$\out$ of possible outcomes, together with the corresponding reduced
state $\varrho_{m|x}$. Given the input letter $x$, the probability of
getting the result $m$ is written as a conditional probability
$p(m|x)$. Then, information is usually understood as ``information in
$m$ about $x$'' and disturbance as ``how well one can undo the state
change $\varrho_x\mapsto\varrho_{m|x}$ for all possible couples
$(x,m)\in\X\times\out$''\footnote{We notice that, since a quantum
  measurement process is defined by several multidimensional
  parameters, i.~e. the source and the measurement apparatus, it is
  extremely demanding to require these latter to be meaningfully
  summarized by means of two positive parameters only,
  i.~e. information and disturbance: the very definition of
  information and disturbance hence represents a major problem in
  itself.}.

While information can be defined to be one of the many known and
basically equivalent measures provided by classical information theory
to quantify the input-output correlations given the joint probability
distribution $p(x,m)$\footnote{Different definitions of information
  gain could however lead, in principle, to different least-disturbing
  measurements. For example, it is known that, for a given ensemble of
  states, the measurement optimizing the mutual information is in
  general different from the measurement optimizing the minimum
  discrimination error: see the analysis of the two-states case in
  Ref.~\cite{access-info}.}, we have (at least) two \emph{in
  principle} inequivalent---still both meaningful---ways to measure
disturbance in a quantum scenario: the first one, adopted e.~g. in
Ref.~\cite{fidelity-tradeoffs}, measures disturbance depending on
``how close'' the states $\varrho_x$ and $\varrho_{m|x}$ are, for all
$x$ and $m$, possibly after a correcting operation performed onto the
output states. We tend to call this way the \emph{average output
  fidelity} approach. The second way, that we call \emph{entanglement
  fidelity} approach, adopted e.~g. in
Ref.~\cite{invert-tradeoffs,nostro}, measures disturbance depending on
``how reversible'', or, equivalently, ``how coherent'' is the
transformation causing the state change, and it is related to the
possibility of reliably transmitting quantum entanglement through the
measurement apparatus. The second approach is generally much more
stringent than the first one, as the following example shows. Let us
consider the simple situation in which the input quantum states
$\varrho_x$ are actually pure orthogonal states. Then, the von Neumann
measurement projecting onto such states is able to fully extract the
information about $x$, without causing any disturbance, when we
understand it in the first sense (indeed, in this case,
$\varrho_{m|x}=\varrho_x$ and $p(x,m)=p(x)\delta_{m,x}$). On the other
hand, a complete projection like a von Neumann measurement causes a
sudden and complete decoherence of any input state onto the
measurement basis: it is hence amongst the most disturbing state
changes possible, if disturbance is understood in the second sense,
since the associated state change completely destroys
entanglement. (We will be more quantitative in the following.)

In fact, there is a third remarkable way to define disturbance, and it
is the one adopted by Ozawa in Ref.~\cite{ozawa-universal}. Ozawa
analyzed the case of a subsequent measurement of two generally non
commuting observables and obtained a universally valid uncertainty
relation, in the spirit of Heisenberg~\cite{heisenberg} and
Robertson~\cite{robertson}, as a formula explicitly involving noise,
disturbance, and pre-measurement uncertainties (i.~e. standard
deviations in the preparation stage). However, we prefer to keep
Ozawa's analysis in a separate class, since he describes the evolution
in the Heisenberg picture, considering observables evolving in time,
while quantum states are fixed. Therefore, the notion of disturbance
Ozawa uses is about observables and cannot be straightforwardly
translated into a notion of disturbance about states, which is the one
usually adopted when dealing with quantum communication scenarios,
like e.~g. quantum cryptography.

The question we address in this paper is: how can we reconcile the
first two mentioned approaches to quantify disturbance, one using
average-output-fidelity--like criteria, the other using
entanglement-fidelity--like criteria? The motivation comes from a
previous work of ours~\cite{nostro} where we derived the closed
balance of information in a quantum measurement process by introducing
a \emph{quantum information gain}, which constitutes an upper bound to
the information that the apparatus is able to extract, independently
of how this information is encoded, and a \emph{quantum disturbance},
which is related to the possibility of deterministically and
coherently undoing the corresponding state change. Such a balance, by
construction, incorporates a tight information-disturbance tradeoff
where disturbance is understood in the second sense. In a way, quantum
information gain and quantum disturbance defined in~\cite{nostro} are
intrinsic features of the measuring instrument, much like the
``technical specs'' of the apparatus, independent of the source that
is to be measured. Moreover, being a \emph{closed} balance, the
formula of Ref.~\cite{nostro} is able to explain the physical causes
for the existence of the tradeoff, the latter being a direct
consequence of the appearance of some hidden correlations between the
input system (undergoing the measurement) and some inaccessible
degrees of freedom (like e.~g. the environment). The approach we
proposed in Ref.~\cite{nostro} hence seems to possess some advantages
with respect to other approaches, at least in terms of compactness and
generality. It has been unclear for a while, however, how to derive
from it a general tradeoff relation also when disturbance is
understood in the first sense, namely, when it is measured in terms of
average output fidelity. The previous example involving orthogonal
states and von Neumann measurements seems to stand as an
insurmountable obstacle to this program. Nevertheless, we will prove
that it is indeed possible to circumvent it by showing that, whenever
the input alphabet is encoded onto non orthogonal pure states (we will
be more precise in the following), then the two notions of disturbance
are equivalent, in the sense that when one is sufficiently small, the other is correspondingly small.
 Such a conclusion basically comes as a corollary of Lemma~3 in
Ref.~\cite{rev-extr}, whose proof we adapted here for our needs. It
also provides the fact that, the more orthogonal the input states get,
the looser the equivalence relation between the two notions of
disturbance becomes, until when, for orthogonal ensembles, the two
definitions become inequivalent, as in the mentioned example. In this
sense, we will have a formula which describes the continuous
transition from the quantum theory, where information extraction
implies disturbance (in whichever way we understand it), to classical
theory, where information can be freely read and copied.

As a drawback of the generality of our analysis (in particular, no
symmetry is assumed for the input ensemble), the mathematical
statement of the main result (Theorem~\ref{theo}) will turn out to be
not very efficient, mainly because of two reasons: the first reason is
that, as we said, we will not assume any sort of symmetry which could
simplify the analysis; the second reason is that we will relate
extensive quantities (like entropy and entropic measures of
information) with non-extensive ones (like fidelity and fidelity-based
measure of disturbance). On the contrary, when considering very
special cases of highly symmetric input ensembles, we will see that,
as one would expect, it is indeed possible to obtain much more compact
formulas. Such a simplification is achieved by considering alternative
entropy-based measures of disturbance which can be more elegantly
related with informational quantities than fidelities. However, a
generalization of this alternative approach seems to be difficult, and
the results we have in this direction are still largely
unsatisfactory. Nonetheless, we present them here in
Section~\ref{sec:ext} as a preliminary step towards further
investigations.

\section{Notation and basic concepts}

In this section we introduce notations and recall basic concepts and
facts that will be useful in the following. Let us consider the
following common situation: some source $\mS$ draws the letter $x$
from an alphabet $\X$, according to the probability distribution
$p(x)$, and correspondingly emits the signal state $\varrho_x^{\s{Q}}$
($\varrho_x^{\s{Q}}\ge 0$ and $\Tr[\varrho_x^{\s{Q}}]=1$) belonging to the
$d$-dimensional Hilbert space $\s{Q}$. (Throughout this paper, all
Hilbert spaces will be finite dimensional.) Such a source is modeled
as a quantum ensemble
\begin{equation}
  \mS:=\{p(x),\varrho_x^{\s{Q}}\}_{x\in\X},
\end{equation}
whose average state is denoted by
$\varrho_\mS^{\s{Q}}:=\sum_xp(x)\varrho_x^{\s{Q}}$. A measurement
$\instr^{\s{Q}}$ is then performed onto the system $\s{Q}$ (the input
of the measurement apparatus) to get information about the encoded
letter $x\in\X$. As proved by Ozawa in Ref.~\cite{ozawa-instr}, the
most general description of any experimentally realizable quantum
measurement is given in terms of a completely positive (CP)
\emph{quantum instrument}, that is, a CP-map--valued measure
$\{\mE_m^{\s{Q}}\}_{m\in\out}$, defined on a set of classical outcomes
$m\in\out$, and normalized such that $\sum_{m\in\out}\mE_m^{\s{Q}}$ is
trace-preserving (i.~e. a channel). When the input state is described
by the density matrix $\varrho^{\s{Q}}$, the probability of obtaining
the outcome $m$ is given by
$p(m|\varrho):=\Tr[\mE_m^{\s{Q}}(\varrho^{\s{Q}})]$; correspondingly,
the state change
$\varrho^{\s{Q}}\mapsto\sigma^{\s{Q}'}_{m|\varrho}:=\mE_m^{\s{Q}}(\varrho^{\s{Q}})/p(m|\varrho)$
occurs, $\s{Q}'$ denoting the output Hilbert space. With a little
abuse of notation, it is also helpful to represent the average state
change caused by the measurement $\instr^{\s{Q}}$ as a black box
performing some fixed operation on the input state, namely, as a
channel with hybrid quantum-classical output system
\begin{equation}\label{eq:instr-channel}
  \instr^{\s{Q}}(\varrho^{\s{Q}}):=\sum_{m\in\out}p(m|\varrho)\sigma^{\s{Q}'}_{m|\varrho}\otimes|m\>\<m|^{\out},
\end{equation}
where $|m\>\<m|^{\out}$ are a set of orthonormal (hence perfectly
distinguishable) vectors in the classical register space $\out$ of
outcomes.

In addition, we will find it useful to introduce an auxiliary
\emph{reference} $\s{R}$ (that we choose isomorphic with $\s{Q}$)
purifying the input state $\varrho^{\s{Q}}$ into $|\Psi^{\s{RQ}}\>$ and
going untouched through the whole measurement process, in such a way
that
\begin{eqnarray}
  |\Psi^{\s{RQ}}\>\mapsto(&\id^{\s{R}}\otimes\instr^{\s{Q}})(\Psi^{\s{RQ}})\nonumber\\
  &=\sum_{m\in\out}p(m|\varrho)\Sigma^{\s{RQ}'}_{m|\varrho}\otimes|m\>\<m|^{\out}\label{eq:ref1},
\end{eqnarray}
where
\begin{equation}\label{eq:ref2}
  \Sigma^{\s{RQ}'}_{m|\varrho}:=\frac{(\id^{\s{R}}\otimes\mE^{\s{Q}}_m)(\Psi^{\s{RQ}})}{p(m|\varrho)}.
\end{equation}
It is clear that
$\Tr_{\s{R}}[\Sigma^{\s{RQ}'}_{m|\varrho}]=\sigma^{\s{Q}'}_{m|\varrho}$, for
all $m\in\out$. Moreover, by denoting
$\tau^{\s{R}}_{m|\varrho}:=\Tr_{\s{Q}'}[\Sigma^{\s{RQ}'}_{m|\varrho}]$, we
have
\begin{equation}\label{eq:ref3}
  \sum_{m\in\out}p(m|\varrho)\tau^{\s{R}}_{m|\varrho}=\Tr_{\s{Q}}[\Psi^{\s{RQ}}]=:\varrho^{\s{R}},
\end{equation}
where $\Psi^{\s{RQ}}$ stands for the projector
$|\Psi^{\s{RQ}}\>\<\Psi^{\s{RQ}}|$. It means that the action of the
instrument $\instr^{\s{Q}}$ on $\s{Q}$ induces the ensemble
decomposition $\{p(m|\varrho),\tau^{\s{R}}_{m|\varrho}\}_{m\in\out}$ on
$\s{R}$.

Even being a formally defined---hence non directly
accessible---system, the reference $\s{R}$ will play a major role in
our analysis. In order to better understand its meaning, it is
suitable to think about $\s{R}$ as the remote system about which we
actually want to extract information by measuring the system $\s{Q}$,
the latter representing in fact just the carrier undergoing the
measurement. Indeed, consider for example the case in which the state
$|\Psi^{\s{RQ}}\>$ is a purification of some state
$\varrho^{\s{Q}}$. Then, all possible sources
$\mS=\{p(x),\varrho_x^{\s{Q}}\}_x$ having average state
$\varrho^{\s{Q}}_\mS$ equal to $\varrho^{\s{Q}}$, namely,
$\varrho^{\s{Q}}=\sum_xp(x)\varrho^{\s{Q}}_x$, are in one-to-one
correspondence with positive-operator--valued measures (POVM) on
$\s{R}$ through the duality relation
\begin{equation}\label{eq:povm-states}
  p(x)\varrho_x^{\s{Q}}=:\Tr_{\s{R}}\left[(R^{\s{R}}_x\otimes\openone^{\s{Q}})\ \Psi^{\s{RQ}}\right].
\end{equation}
In other words, when thinking of a classical-to-quantum communication
scenario, the system $\s{Q}$ represents the receiver, the source $\mS$
models the quantum channel $x\mapsto\varrho^{\s{Q}}_x$ along which the
classical information is sent, while the reference system $\s{R}$ can
be understood as the sender's system which retains the information
about the encoded index $x$.

\section{Information gains and disturbances}

\subsection{Information gains: mutual information, accessible
  information, and quantum information gain}

Let us suppose that we know exactly the source $\mS$, in the sense
that we know the probability distribution $p(x)$ and the states
$\varrho_x$ which are associated with the letters $x\in\X$. In this
case, given the instrument $\instr^{\s{Q}}$, it is straightforward to
compute the joint probability distribution
$p(m,x)=\Tr[\mE_m(\varrho_x)]$, and, from it, the correlations between
the input letter and the measurement readout. We choose here to
measure them as classical mutual information
\begin{equation}\label{eq:classical_mutual_info}
I(\X:\out):=\sum_{x\in\X}\sum_{m\in\out}p(x,m)\log_2\frac{p(x,m)}{p(x)p(m)},
\end{equation}
where we implicitly used the Bayes' rule $p(x,m)=p(x)p(m|x)$.

Imagine now that we ignore the exact structure of the source $\mS$,
but its average state $\varrho_\mS$. Even in this case we can define a
mutual information--like quantity that describes how much information
the instrument is able to extract: this quantity is the
\emph{accessible information}, when the measurement is fixed, and we
maximize over all possible ensemble
$\mS'=\{p'(x'),\varrho'_{x'}\}_{x'\in\X'}$ with average state
$\varrho_{\mS'}=\varrho_\mS$, that is
\begin{equation}
  \Iacc(\varrho_\mS,\instr):=\max_{\mS':\varrho_{\mS'}=\varrho_\mS}I(\X':\out).
\end{equation}
This quantity is well-known to be extremely hard to compute
explicitly. It is then useful to introduce another quantity, which is
much easier to calculate, yet equivalent\footnote{Via bounds that,
  being dimension-dependent, must be handled with care: see the
  following.} to $\Iacc$ for finite dimensional systems. In
Ref.~\cite{nostro} , we introduced the \emph{quantum} information gain
defined as the entropic defect~\cite{chi} of the ensemble induced on
the reference $\s{R}$ by the action of the instrument on $\s{Q}$
\begin{equation}\label{eq:qinfo}
  \iota(\varrho_\mS,\instr):=S(\varrho^{\s{R}})-\sum_{m\in\out}p(m|\varrho_\mS)S(\tau^{\s{R}}_{m|\varrho_S}),
\end{equation}
where $S(\varrho):=-\Tr[\varrho\log_2\varrho]$ is the von Neumann entropy, and
other notations follows Eqs.~(\ref{eq:ref1})-(\ref{eq:ref3}). This
quantity is what we are searching for, in that it is easy to compute
and in Appendix~A we prove that
\begin{equation}\label{eq:info-equiv}
  \Iacc(\varrho_\mS,\instr)\le\iota(\varrho_\mS,\instr)\le\mathrm{t}\left((2d-1)\sqrt{2\Iacc(\varrho_\mS,\instr)}\right),
\end{equation}
where $d:=\dim\s{Q}$ and $\mathrm{t}(x)$ is a positive, continuous,
monotonic increasing function such that $\mathrm{t}(0)=0$. (The first
inequality is the Holevo bound on accessible
information~\cite{holevo}.) In other words, for finite dimensional
systems, $\Iacc(\varrho_\mS,\instr)\to 0$ if and only if
$\iota(\varrho_\mS,\instr)\to 0$. About this point, as we mentioned
before, an important \emph{caveat} is in order: the appearance of the
dimension $d$ in Eq.~(\ref{eq:info-equiv}) makes the equivalence
relation quite weak as dimension increases. Such a factor is
reminiscent of the phenomenon known as \emph{locking} of classical
correlations~\cite{michal-IBM}, and it is in general
unavoidable~\cite{bariska}.

As a concluding remark, notice that, even if the reference system
$\s{R}$ explicitly appears in the definition~(\ref{eq:qinfo}), quantum
information gain only depends on the average input state
$\varrho_\mS^{\s{Q}}$ and on the instrument $\instr^{\s{Q}}$,
regardless of the particular purification $|\Psi^{\s{RQ}}\>$ chosen.

\subsection{Disturbances: average output fidelity, entanglement
  fidelity, and quantum disturbance}

Let us suppose that we know exactly the source $\mS$, in the sense
that we know the probability distribution $p(x)$ and the states
$\varrho_x$ which are associated with the letters $x\in\X$. Moreover,
let us suppose that we know exactly the quantum instrument
$\instr^{\s{Q}}$, in the sense that we know the outcome set $\out$ and
the set of associated CP-maps $\{\mE^{\s{Q}}_m\}_{m\in\out}$. Having
access to the measurement readout $m$, but not to the input letter $x$
being unknown to the receiver, the best (deterministic) correction we
could try in order to reduce state disturbance, is in principle to
engineer, depending on the source and on the instrument, a set of
correcting channels $\{\mR_m^{\s{Q}'}\}_{m\in\out}$, mapping $\s{Q}'$
back to $\s{Q}$, and to apply them onto the output state according to
the measurement readout $m$. The maximum output fidelity approach aims
at finding channels $\mR_m^{\s{Q}'}$ achieving the quantity
\begin{equation}\label{eq:out-fid}
  \max_{\mR_m}\sum_xp(x)\fid\left(\frac{\mR_m^{\s{Q}'}(\mE^{\s{Q}}_m(\varrho^{\s{Q}}_x))}{p(m|x)},\varrho^{\s{Q}}_x\right),
\end{equation}
for all $m\in\out$, where
$\fid(\varrho,\sigma):=(\Tr|\sqrt{\varrho}\sqrt{\sigma}|)^2$ is the
fidelity. We however relax this condition by considering a situation
where we are content to see that just the average value
\begin{equation}\label{eq:av-fid}
  \cfid(\mS,\instr):=\max_{\{\mR_m\}}\sum_{x\in\X}p(x)\fid\left(\sum_{m\in\out}\mR_m^{\s{Q}'}(\mE^{\s{Q}}_m(\varrho^{\s{Q}}_x)),\varrho^{\s{Q}}_x\right)
\end{equation}
is close to one. This means that, whenever $\cfid<1$, there exists at
least one state $\varrho_x$ which gets irrecoverably disturbed by the
corrected channel $\sum_m\mR_m\circ\mE_m$. In turns, due to concavity
of fidelity, this means that, for such state $\varrho_x$, there exists
at least one outcome $m$ for which the corrected
fidelity~(\ref{eq:out-fid}) is strictly less than one.

Suppose now that we instead \emph{ignore} the exact structure of the
source $\mS$, but know its average state $\varrho_\mS$ only. In this
case, the entanglement fidelity~\cite{schum} approach provides a
suitable definition for disturbance. The quantity to consider in this
case is
\begin{equation}
  \max_{\mR_m}\fid\left(\frac{(\id^{\s{R}}\otimes\mR_m^{\s{Q}'}\circ\mE^{\s{Q}}_m)(\Psi^{\s{RQ}})}{p(m|\rho_\mS)},\Psi^{\s{RQ}}\right)
\end{equation}
for all possible outcomes $m\in\out$, where, as usual,
$|\Psi^{\s{RQ}}\>$ is a purification of the average state
$\varrho_\mS^{\s{Q}}$. As before, we relax a bit the request, also to
simplify the notation, and focus on the average quantity
\begin{equation}\label{eq:ent-fid}
\cFe(\varrho_{\mS},\instr):=\max_{\{\mR_m\}}\fid\left(\sum_{m\in\out}(\id^{\s{R}}\otimes\mR_m^{\s{Q}'}\circ\mE^{\s{Q}}_m)(\Psi^{\s{RQ}}),\Psi^{\s{RQ}}\right).
\end{equation}
Incidentally, notice that, in this setting, the restoring channels
$\mR^{\s{Q}'}_m$ cannot depend on the ensemble $\mS$, but on its
average state $\varrho_\mS$ only. A value close to one for
$\cFe(\varrho_\mS,\instr)$ means that, on the support of $\varrho_\mS$, the
channel $\sum_{m\in\out}\mR_m^{\s{Q}'}\circ\mE_m^{\s{Q}}$ is close to
the identity channel $\id^{\s{Q}}$. This in turn implies that, not
only states are transmitted with high output fidelity, but also
entanglement can be reliably sent. Entanglement fidelity is known to
provide a lower bound on the worst possible average output fidelity,
in the sense that
\begin{equation}\label{eq:av-fid-lower-bound}
  \cFe(\varrho_{\mS},\instr)\le\min_{\mS':\varrho_{\mS'}=\varrho_\mS}\cfid(\mS',\instr).
\end{equation}
This is the reason why the entanglement fidelity criterion is
generally much more severe than the average output fidelity
criterion. The example provided in the introduction exhibits the
extreme case in which the two criteria are strictly
inequivalent. Notice that a value $\cFe<1$ means that there is at
least one outcome $m$ for which the corresponding state change map is
irreversible on the support of $\varrho_\mS$.

The quantum disturbance, defined in Ref.~\cite{nostro} as
\begin{equation}\label{eq:q-dist}
  \delta(\varrho_\mS,\instr):=S(\varrho^{\s{Q}}_\mS)-I_c^{\s{R}\to\s{Q}'\out}\left((\id^{\s{R}}\otimes\instr^{\s{Q}})(\Psi^{\s{RQ}})\right),
\end{equation}
where
$I_c^{\s{A}\to\s{B}}(\varrho^{\s{AB}}):=S(\varrho^{\s{B}})-S(\varrho^{\s{AB}})$
is the \emph{coherent information} from subsystem $\s{A}$ to subsystem
$\s{B}$ for the bipartite state $\varrho^{\s{AB}}$, is a quantity bounded
between zero and $S(\varrho^{\s{Q}}_\mS)$ and it is equivalent to the
entanglement fidelity~(\ref{eq:ent-fid}), in the sense
that~\cite{nostro}
\begin{equation}\label{eq:entropic-tradeoff}
\frac 14\left(1-\cFe(\varrho_{\mS},\instr)\right)^2\le\delta(\varrho_\mS,\instr)\le\textrm{g}(1-\cFe(\varrho_\mS,\instr)),
\end{equation}
where $\textrm{g}(x)$ is some positive, continuous,
monotonic increasing function such that $\textrm{g}(0)=0$. Moreover,
the tradeoff relation is always obeyed~\cite{nostro}
\begin{equation}\label{eq:18}
  \iota(\varrho_\mS,\instr)\le\delta(\varrho_\mS,\instr),
\end{equation}
where $\iota(\varrho_\mS,\instr)$ is defined in~(\ref{eq:qinfo}). The
equality in~(\ref{eq:18}) is achieved by ``single-Kraus'' or
``multiplicity free'' instruments, that are those for which every map
$\mE_m$ defining the instrument $\instr$ is represented by a single
contraction, that is, $\mE_m(\varrho)=E_m\varrho E_m^\dag$. In this
sense, single-Kraus instruments can be considered as least-disturbing
measurements. At this point, the question remains open whether or not
quantum disturbance $\delta(\varrho_\mS,\instr)$ is equivalent to
average output fidelity $\cfid(\mS,\instr)$ as well, and, if it is, to
which extent---in fact, they \emph{cannot} be always equivalent, as
the von Neumann measurement example, for which $\cfid(\mS,\instr)=1$
while $\delta(\varrho_\mS,\instr)=S(\varrho_\mS)$, proves. To answer
this question will be the aim of the next section.

\section{Unifying disturbances for irreducible ensembles of pure
  states}

Let us suppose now that the source $\mS$ produces pure states, that
is, $\varrho^{\s{Q}}_x=\psi_x^{\s{Q}}$, for all $x\in\X$. Now, let us
consider all possible \emph{$N$-complete paths}, namely, all possible
$N$-sequences $\x^N=(x_1,x_2,\cdots,x_N)\in\X^{\times N}$ such that
the corresponding set of states
\begin{equation}
  \left\{|\psi^{\s{Q}}_{x_1}\>,|\psi^{\s{Q}}_{x_2}\>,\cdots,|\psi^{\s{Q}}_{x_N}\>\right\}
\end{equation}
contains, at least once, every state $|\psi^{\s{Q}}_x\>$ emitted by
the source\footnote{We could relax this condition by asking valid
  complete paths to form spanning sets for
  $\supp(\varrho_\mS^{\s{Q}})$. This technicality, however, even being
  useful is some cases, would force our analysis to be notationally
  heavy. We hence prefer to loose a bit in the efficiency of our
  bounds while gaining in clarity.}. Let us denote the set of all
possible $N$-complete path for a given source $\mS$ as
$\mathscr{P}_\mS^N$, and let
$\mathscr{P}_\mS=\bigcup_N\mathscr{P}_\mS^N$ be the set of all
possible complete paths. Let us moreover define the following function
of a given ensemble $\mS$
\begin{equation}\label{eq:eta}
  \eta(\mS):=\max_{\x^N\in\mathscr{P}_\mS}\frac{\min_{1\le i\le N-1}|\<\psi^{\s{Q}}_{x_i}|\psi^{\s{Q}}_{x_{i+1}}\>|}{N},
\end{equation}
where the maximum is taken over all possible complete paths. Following
Ref.~\cite{rev-extr}, we say that the source $\mS$ is
\emph{irreducible} if and only if $\eta(\mS)>0$, that is equivalent to
say that the states $|\psi^{\s{Q}}_i\>$ cannot be divided into two or
more orthogonal subsets. The notion of irreducibility makes rigorous
the notion of ``non-orthogonal ensembles'' used in the introduction:
in fact, within a given ensemble of pure states, some of them could
well be orthogonal, yet the ensemble being irreducible. Notice
moreover that, implicit in Eq.~(\ref{eq:eta}) there is another
tradeoff: we can make the numerator larger by considering ``more
complicated'' paths, but, eventually, also the denominator gets
larger. Hence, to find the optimum value of $\eta(\mS)$ is a tricky
procedure. However, for sources emitting a finite number of states,
the solution (maybe corresponding to non-unique choices of optimal
complete paths) always exists.

The quantity $\eta(\mS)$ is a good indicator of the degree of
irreducibility, or quantumness, of the ensemble $\mS$. There is
however a hidden subtlety: till now we did not consider the role of
the probability distribution $p(x)$ of the source yet. Imagine, for
example, that a source is irreducible, in the sense that
$\eta(\mS)>0$, but only because of one state $|\psi_\emptyset\>$ being
non-orthogonal with any other, whose probability of occurrence
$p(\emptyset)$ is however \emph{very} small. Such a source is then
irreducible, but only ``weakly'' irreducible: we should take this
possibility into account. We then define the function
\begin{equation}\label{eq:zeta}
\zeta(\mS):=\eta(\mS)\cdot\left[\min_xp(x)\right].
\end{equation}
Again, since by definition $\min_xp(x)>0$, the source of pure states
$\mS$ is irreducible is and only if $\zeta(\mS)>0$, but now,
contrarily to what happens for $\eta(\mS)$, we are automatically
weighing the role of the probability distribution $p(x)$. In the
following theorem we see how $\zeta(\mS)$ enters in our discussion:

\begin{theo}\label{theo}
  Let $\mS=\{p(x),|\psi^{\s{Q}}_x\>\}_{x\in\X}$ be an ensemble of pure
  states, with average state $\varrho_\mS$, undergoing the quantum
  measurement described by the instrument $\instr^{\s{Q}}$. Then
\begin{eqnarray}
  \frac 14\left(1-\cfid(\mS,\instr)\right)^2&\le\frac 14\left(1-\cFe(\varrho_\mS,\instr)\right)^2\label{eq:theo1a}\\
&\le\delta(\varrho_\mS,\instr)\label{eq:theo1b}\\
&\le\textrm{\emph{f}}\left(\frac{\sqrt{1-\cfid(\mS,\instr)}}{\zeta(\mS)}\right),\label{eq:theo1c}
\end{eqnarray}
where $\textrm{\emph{f}}(x)$ is some positive, continuous,
monotonic increasing function such that
$\textrm{\emph{f}}(0)=0$.\qquad$\square$
\end{theo}
In other words, whenever the ensemble $\mS$ is irreducible, namely,
$\zeta(\mS)>0$, then the notions of disturbance originating from
average output fidelity, from entanglement fidelity, and from quantum
disturbance are all equivalent, in the sense that
$\cfid(\mS,\instr)\to1 \Leftrightarrow \cFe(\varrho_{\mS},\instr)\to1
\Leftrightarrow \delta(\varrho_\mS,\instr)\to 0$, and, equivalently,
$\cfid(\mS,\instr)<1 \Leftrightarrow \cFe(\varrho_{\mS},\instr)<1
\Leftrightarrow \delta(\varrho_\mS,\instr)>0$. On the other hand, as
$\zeta(\mS)\to 0$ (a situation happening when $\eta(\mS)\to 0$ or
$\min_xp(x)\to 0$), while a high entanglement fidelity always implies
a high average output fidelity, the converse direction becomes weaker
and weaker\footnote{As noted in a previous footnote, by allowing a
  generalized notion of complete paths, we could also get rid here of
  signals---states with their corresponding probabilities---which are
  not necessary to span the subspace $\supp(\rho_\mS)$, and possibly
  obtain an effective parameter $\tilde\zeta(\mS)$ which is larger
  than the one defined in Eq.~(\ref{eq:zeta}). For the sake of
  clarity, however, we will leave this improvement as an exercise for
  the interested reader.}. The example we mentioned in the
introduction (involving orthogonal states and von Neumann measurement)
where the two disturbances are irreconcilably inequivalent, even
looking somehow over-simplified, actually captures all the essential
features of those situations where average output fidelity and
entanglement fidelity are really inequivalent quantities.

{\bf Proof of Theorem~\ref{theo}}. Inequalities~(\ref{eq:theo1a})
and~(\ref{eq:theo1b}) follow directly from
Eqs.~(\ref{eq:av-fid-lower-bound})
and~(\ref{eq:entropic-tradeoff}). The last
inequality~(\ref{eq:theo1c}) is a consequence of the result proved in
the Appendix~B, which in turn follows the proof of Lemma~3 in
Ref.~\cite{rev-extr}. In fact, the quantum disturbance
$\delta(\varrho_\mS,\instr)$ is defined as the coherent information
loss of the channel $\instr^{\s{Q}}$, as it is defined in
Eq.~(\ref{eq:instr-channel}), mapping the input system $\s{Q}$ in the
output quantum-classical system $\s{Q}'\out$. Moreover, whatever
channel $\mR^{\s{Q}'\out}$, applied after $\instr^{\s{Q}}$ in order to
map the hybrid output system $\s{Q}'\out$ back to the input system
$\s{Q}$, returns in fact a set of channels
$\{\mR^{\s{Q}'}_m\}_{m\in\out}$, each of them applied conditionally on
the readout $m$, all of them mapping $\s{Q}'$ back to $\s{Q}$. This is
indeed the way in which we defined $\cfid(\mS,\instr)$ in
Eq.~(\ref{eq:av-fid}). Therefore, we can straightforwardly apply the
statement of Proposition~1 in Appendix~B and get
Eq.~(\ref{eq:theo1c}).\qquad$\blacksquare$

\section{Entropy-based notion of average disturbance}\label{sec:ext}

The main content of Theorem~\ref{theo} is to make explicit the
relation existing between disturbances defined in terms of average
output fidelity and entanglement fidelity. Quantum disturbance appears
there because of Eq.~(\ref{eq:entropic-tradeoff}), which states that
quantum disturbance is essentially equivalent to
entanglement-fidelity--based disturbance. It is now reasonable to ask
whether one can devise an entropy-based analog also for
average-output-fidelity--based disturbance, in the hope that the
relation existing between such a new measure of disturbance and
quantum disturbance can be expressed in a form simpler than
Eqs.~(\ref{eq:theo1a})-(\ref{eq:theo1c}).

Let us consider an input ensemble $\mS:=\{p(x),\varrho_x\}_{x\in\X}$
undergoing a channel $\mE$. It is known that its entropy defect
$\chi(\mS):=S(\varrho_\mS)-\sum_xp(x)S(\varrho_x)$ behaves
monotonically under the action of a channel, that is, the entropy
defect loss
\begin{equation}
\Delta\chi(\mS,\mE):=\chi(\mS)-\chi(\mE(\mS))
\end{equation}
is always nonnegative. Moreover (see Ref.~\cite{michal} and
Lemma~\ref{lemma:2} in \ref{app:b}) it is known that the condition
$\cfid(\mS,\mE)\to 1$ implies that $\Delta\chi(\mS,\mE)\to 0$
correspondingly. On the other hand, the opposite direction is known to
hold only in the exact case, that is, $\Delta\chi(\mS,\mE)=0$ implies
that $\cfid(\mS,\mE)=1$ (see for example~\cite{strong-subadd}). In
other words, at the moment we cannot state that entropy defect loss
and corrected average output fidelity are truly equivalent measures of
disturbance, since we do not know whether a small entropy defect loss
implies a correspondingly small average fidelity loss.

Anyway, let us postulate for the moment that entropy defect loss
$\Delta\chi(\mS,\mE)=0$ is the sought informational analog of average
output fidelity, much like quantum disturbance represents the
informational analog of entanglement fidelity. In the case of a
quantum measurement process, when we consider the ``channelized''
action of the instrument $\instr$ as given in
Eq.~(\ref{eq:instr-channel}), it is straightforward to compute
\begin{equation}
\chi(\instr(\mS))=I(\X:\out)+\sum_mp(m)\chi(\overline{\mS}_m),
\end{equation}
where $I(\X:\out)$ represents the (classical) mutual information
between the input letter and the measurement outcome, as defined in
Eq.~(\ref{eq:classical_mutual_info}), and
\begin{equation}
  \overline{\mS}_m:=\left\{p(x|m),\frac{\mE_m(\varrho_x)}{p(m|x)}\right\}_{x\in\X}
\end{equation}
is the ensemble output by the measuring apparatus, given the outcome
readout $m$. Notice that, for every conditional output ensemble
$\overline{\mS}_m$, the average state $\sigma_{\overline{\mS}_m}$ is
\begin{equation}
  \sigma_{\overline{\mS}_m}:=\sum_{x\in\X}p(x|m)\frac{\mE_m(\varrho_x)}{p(m|x)}=\frac{\mE_m(\varrho_\mS)}{p(m)},
\end{equation}
according to Bayes' rule. Then, the entropy defect loss is equal to
\begin{eqnarray}
  \Delta\chi(\mS,\instr):&=\chi(\mS)-\chi(\instr(\mS))\nonumber\\
  &=\sum_mp(m)\left[\chi(\mS)-\chi(\overline{\mS}_m)\right]-I(\X:\out).
\end{eqnarray}

Let us now focus on a very special case of input ensemble (introduced
in Ref.~\cite{CW} and generalized in Ref.~\cite{shredder}). It is
possible to define such an ensemble for any given ensemble average
state $\varrho_\mS$ as follows: given the average state $\varrho_\mS$
on system $\s{Q}$, after having purified it as $|\Psi^{\s{RQ}}\>$ by
means of a reference system $\s{R}$ isomorphic to $\s{Q}$, one defines
the pure states $|\psi_x^{\s{Q}}\>$ along with their occurrence
probabilities $p(x)$ as
\begin{equation}
  p(x)\psi_x^{\s{Q}}:=\Tr_{\s{R}}[(\phi_x^{\s{R}}\otimes\openone^{\s{Q}})\ \Psi^{\s{RQ}}].
\end{equation}
In the above equation, the vectors $\{\phi_x^{\s{R}}\}_{x\in\X}$
constitute a rank-one POVM on $\s{R}$ with $2d$ elements, such that
$\Tr[\phi_x]=1/2d$ for all $x$, and defined as follows: $d$ of them
form a basis diagonalizing $\Tr_{\s{Q}}[\Psi^{\s{RQ}}]$, while the
other $d$ vectors form another basis, mutually unbiased with respect
to the first one.

For this particular input ensemble of pure states, that we call of
Christandl-Winter type, it is known that~\cite{CW,shredder}
\begin{equation}
  \Delta\chi(\mS,\instr)\le\delta(\varrho_\mS,\instr)\le 2\Delta\chi(\mS,\instr),
\end{equation}
that is, entropy defect loss and quantum disturbance are equivalent
quantities, and their relation is strikingly more stringent than the
relation existing between average output fidelity and entanglement
fidelity as given in Theorem~\ref{theo}. However, as we stated in the
introduction, the extension of this approach to the more general case
where the input ensemble does not come from such a highly symmetric
construction seems to be a hard path to pursue, yet worth further
investigations.

\section{Conclusion and discussion}

With Theorem~1 we showed that, whenever the input source is
sufficiently rich in its structure---i.~e. it is irreducible---the
vast majority of definitions of disturbance, introduced in the
literature so far, actually turn out to be equivalent. In particular,
we focused on the relations existing between average output fidelity
and entanglement fidelity, without imposing any symmetry on the source
distribution. Our analysis could hence unify different approaches to
the quantification of information-disturbance tradeoffs in quantum
theory. The general bounds obtained here exploit Fannes-like bounds on
entropic quantities, and, for this reason, relate extensive quantities
with non-extensive ones. In the very particular case of input
ensembles of the Christandl-Winter type, we showed a way to circumvent
this difficulty and obtain much more stringent bounds. A future
research direction is to extend this latter approach to more general
situations.

\ack F.~B. acknowledges Japan Science and Technology Agency for
support through the ERATO-SORST Quantum Computation and Information
Project.  M.~H. is supported by EC IP SCALA. Part of this work was
done while F.~B. was visiting the Institute of Theoretical Physics and
Astrophysics at the University of Gda\'nsk.

\appendix\setcounter{section}{0}

\section{Accessible information and quantum information gain are
  equivalent criteria to measure information gain}

We exploit here a method introduced in Ref.~\cite{erasure}. Let us
consider an ensemble $\mS=\{p(x),\varrho_x\}_{x\in\X}$. The accessible
information is defined as the maximum over all possible POVM's of the
mutual information\footnote{In fact, in the text we came to accessible
  information from the dual point of view, where the measurement
  (i.~e. the POVM) is fixed and the maximization is done with respect
  to the input ensemble: thanks to the POVM/ensembles duality
  relation~(\ref{eq:povm-states}) the two approaches are equivalent.}
\begin{equation}
  \Iacc(\mS):=\max_{\{P_m\}_{m\in\out}}I(\X:\out).
\end{equation}
Now, let $\chi(\mS)$ be the entropy defect of the ensemble
$\mS$~\cite{chi}
\begin{equation}\label{eq:chi}
\chi(\mS):=S(\varrho_\mS)-\sum_xp(x)S(\varrho_x),
\end{equation}
where $S(\varrho):=-\Tr[\varrho\log_2\varrho]$ is the von Neumann entropy. Holevo bound states that~\cite{holevo}
\begin{equation}
  \Iacc(\mS)\le\chi(\mS).
\end{equation}
By rearranging terms in Eq.~(\ref{eq:chi}), we can write
\begin{equation}
  \chi(\mS)=\sum_xp(x)(S(\varrho_\mS)-S(\varrho_x)).
\end{equation}
By Fannes inequality~\cite{hayashi}, we know that
\begin{eqnarray}
  \chi(\mS)&=\sum_xp(x)(S(\varrho_\mS)-S(\varrho_x))\nonumber\\
  &\le\sum_xp(x)|S(\varrho_\mS)-S(\varrho_x)|\nonumber\\
  &\le\sum_xp(x)\mathrm{t}\left(\N{\varrho_x-\varrho_\mS}_1\right)\nonumber\\
  &\le\mathrm{t}\left(\sum_xp(x)\N{\varrho_x-\varrho_\mS}_1\right)\label{eq:dove},
\end{eqnarray}
where $\mathrm{t}(x)$ is a positive, continuous, monotonic increasing,
concave function such that $\mathrm{t}(0)=0$ and $\N{X}_1:=\Tr|X|$ is
the trace-norm. In particular, for $x\le 1$, we can take
\begin{equation}
\mathrm{t}(x)=x\log_2\frac{2\sqrt{d-1}}{x},
\end{equation}
where $d$ is the dimension of the
underlying space~\cite{audenaert}.

Let us now make a small detour, and introduce so-called
\emph{informationally complete POVM's}, namely, those POVM's whose
elements also form an operator basis. This means that, being
$\{P_m\}_m$ an info-complete POVM, there exists a \emph{dual frame}
$\{K_m\}_m$, with Hermitian operators $K_m$, such that
\begin{equation}\label{eq:expansion}
X=\sum_m\Tr[XP_m]K_m,
\end{equation}
for all operators $X$. We consider here a particular info-complete
POVM, that is, $\{P_g\}_{g\in\boldsymbol{G}}$ defined as
\begin{equation}\label{eq:specialized-povm}
P_g:=\frac 1dU_g\phi U_g^\dag,
\end{equation}
where $U_g$ is a unitary representation of the group $\mathbb{SU}(d)$,
and $\phi$ is a $d$-dimensional pure state. In Ref.~\cite{perinotti}
the canonical dual frame has been explicitly calculated, and it holds that
\begin{equation}
  \N{K_g}_1=2d-1,\qquad\forall g.
\end{equation}

Let us go back to out original aim, i.~e. to give an upper bound to
$\chi(\mS)$, and continue from Eq.~(\ref{eq:dove})
\begin{eqnarray}
  \chi(\mS)&\le\mathrm{t}\left(\sum_xp(x)\N{\varrho_x-\varrho_\mS}_1\right)\nonumber\\
  &=\mathrm{t}\left(\sum_xp(x)\N{\sum_gp(g|x)K_g-p(g)K_g}_1\right)\label{dove0}\\
  &\le\mathrm{t}\left((2d-1)\sum_xp(x)\sum_g\N{p(g|x)-p(g)}_1\right)\nonumber\\
  &\le\mathrm{t}\left((2d-1)\sqrt{2I(\X:\boldsymbol{G})}\right)\label{dove1}\\
  &\le\mathrm{t}\left((2d-1)\sqrt{2\Iacc(\mS)}\right).\label{dove2}
\end{eqnarray}
Eq.~(\ref{dove0}) uses the expansion formula~(\ref{eq:expansion})
specialized to the info-complete POVM~(\ref{eq:specialized-povm}),
while to obtain Eq.~(\ref{dove1}) we used Pinsker inequality
\begin{equation}
\N{\varrho-\sigma}_1^2\le2D(\varrho\|\sigma),
\end{equation}
where
$D(\varrho\|\sigma):=\Tr[\varrho\log_2\varrho-\varrho\log_2\sigma]$ is
the \emph{quantum relative entropy}. Eq.~(\ref{dove2}) comes from the
very definition of accessible information, which is defined as the
maximum over all possible POVM's.

In conclusion, we have that
\begin{equation}
\Iacc(\mS)\le\chi(\mS)\le\mathrm{t}\left((2d-1)\sqrt{2\Iacc(\mS)}\right),
\end{equation}
that means that, for finite dimensional systems, $\Iacc(\mS)\to 0$ if
and only if $\chi(\mS)\to 0$. In other words, accessible information
and entropy defect are equivalent quantities.

Notice that in passing by we actually proved that
\begin{equation}
  \sum_xp(x)\N{\varrho_x-\varrho_\mS}_1\le(2d-1)\sqrt{2\Iacc(\mS)}.
\end{equation}
Such a bound is a little better than the one proved, with different
techniques, in Ref.~\cite{michal-IBM}.

\section{Average output fidelity and entanglement fidelity are
  equivalent criteria to measure disturbance for irreducible pure
  states ensembles}\label{app:b}

In the following, let us consider a channel, that is, a completely
positive trace-preserving map $\mE:\s{Q}\to\s{Q}'$, and a state
$\varrho$ which undergoes $\mE$. As it is often done, we proceed here
with the tripartite purification of the whole setting. The first step
is to purify the input state $\varrho^{\s{Q}}$ into a bipartite pure
state $|\Psi^{\s{RQ}}\>$, shared with a reference system $\s{R}$, as
we did in Section~2. The second step consists in constructing the
isometry $V:\s{Q}\to\s{Q}'\otimes\s{A}$ (the isometric condition reads
$V^\dag V=\openone^{\s{Q}}$) such that
\begin{equation}\label{eq:stine}
\mE(\varrho)=\Tr_{\s{A}}[V\varrho V^\dag],
\end{equation}
for all input state $\varrho$. The auxiliary system $\s{A}$ is sometimes
called \emph{ancilla} (or environment) and the construction in
Eq.~(\ref{eq:stine}) is usually referred to as the Stinespring's
dilation of $\mE$. Moreover, Stinespring's dilation induces an
ancillary channel $\tilde\mE:\s{Q}\to\s{A}$ defined as
\begin{equation}
\tilde\mE(\varrho):=\Tr_{\s{Q}'}[V\varrho V^\dag],
\end{equation}
for all $\varrho$. The channel $\tilde\mE$ is defined up to local
isometric transformations on $\s{A}$. By combining these two
``purifications'', we can write the tripartite pure state
\begin{equation}
  |\Upsilon^{\s{RQ}'\s{A}}\>:=(\openone^{\s{R}}\otimes V^{\s{Q}})|\Psi^{\s{RQ}}\>,
\end{equation}
such that $\Tr_{\s{RA}}[\Upsilon^{\s{RQ}'\s{A}}]=\mE(\varrho)$.

For our later convenience, we now introduce two useful notation. Given
an ensemble $\mS:=\{p(x),\varrho_x\}_x$, we denote with $\chi(\mS)$ its
entropy defect, in formula
\begin{equation}
\chi(\mS):=S(\varrho_\mS)-\sum_xp(x)S(\varrho_x),
\end{equation}
where $S(\varrho):=-\Tr[\varrho\log_2\varrho]$ is the von Neumann
entropy. By the monotonicity property of quantum relative entropy,
after the action of a channel, the entropy defect drops, and we denote
the difference between input $\chi$ and output $\chi$,
\begin{equation}
  \Delta\chi(\mS,\mE):=\chi(\mS)-\chi(\mE(\mS)).
\end{equation}
We also define the coherent information loss $\delta(\varrho,\mE)$, due
to the action of a channel $\mE^{\s{Q}}:\s{Q}\to\s{Q}'$ onto a state
$\varrho^{\s{Q}}$ which is a subsystem of a larger entangled pure state
$|\Psi^{\s{RQ}}\>$ as
\begin{equation}
\delta(\varrho,\mE):=S(\varrho^{\s{Q}})-I_c^{\s{R}\to\s{Q}'}\left((\id^{\s{R}}\otimes\mE^{\s{Q}})(\Psi^{\s{RQ}})\right),
\end{equation}
where
$I_c^{\s{A}\to\s{B}}(\varrho^{\s{AB}}):=S(\varrho^{\s{B}})-S(\varrho^{\s{AB}})$
is the \emph{coherent information} from subsystem $\s{A}$ to subsystem
$\s{B}$ for the bipartite state $\varrho^{\s{AB}}$. We can now state the
following


\begin{lemma}
  Let $\mE:\s{Q}\to\s{Q}'$ be a channel. Let
  $\mS:=\{p(x),|\psi_x\>\}_x$ be an input ensemble of pure states. Then,
\begin{equation}\label{eq:uno}
  \delta(\varrho_\mS,\mE)=\Delta\chi(\mS,\mE)+\chi(\tilde\mE(\mS)),
\end{equation}
that is, coherent information loss equals the sum of input ensemble
entropy defect loss plus the entropy defect of the induced ancillary
ensemble.\qquad$\square$
\end{lemma}
{\bf Proof}. The proof simply follows by direct
inspection, as done in Ref.~\cite{horo-lloyd-winter}. Notice that the
condition of having pure input states is
crucial.\qquad$\blacksquare$\bigskip

Now, analogously to Eq.~(\ref{eq:av-fid}), let us define the corrected
average output fidelity for a channel $\mE:\s{Q}\to\s{Q}'$ (there are
no classical outcome here, only quantum, as the former is a special
case of the latter) when the input ensemble is
$\mS:=\{p(x),\varrho_x\}_{x\in\X}$ as
\begin{equation}
  \cfid(\mS,\mE):=\max_{\mR}\sum_{x\in\X}p(x)\fid\Big(\mR(\mE(\varrho_x)),\varrho_x\Big),
\end{equation}
where the maximum is taken over all possible channels $\mR$ mapping
$\s{Q}'$ back to $\s{Q}$.


\begin{lemma}\label{lemma:2}
We have that
\begin{equation}\label{eq:f1}
  \Delta\chi(\mS,\mE)\le\mathrm{f}_1(1-\cfid(\mS,\mE)),
\end{equation}
where $\mathrm{f}_1(x)$ is a positive, continuous, monotonic
increasing, concave function such that
$\mathrm{f}_1(0)=0$.\qquad$\square$
\end{lemma}
{\bf Proof}. Following Ref.~\cite{michal}, given two sources of $K$
states $\mS:=\{p(x),\varrho_x\}_x$ and $\mS':=\{p(x),\varrho_x'\}_x$,
we can apply the refined Fannes' continuity relation for von Neumann
entropy~\cite{audenaert} and get
\begin{equation}\label{eq:due}
  |\chi(\mS)-\chi(\mS')|\le 2K\sqrt{\epsilon}\log_2\frac{d_{\s{Q}}}{\epsilon},
\end{equation}
where $\epsilon:=1-\sum_xp(x)\fid(\rho_x,\rho_x')$, for the fidelity
defined as $\fid(\rho,\sigma):=
(\Tr|\sqrt{\rho}\sqrt{\sigma}|)^2$. The statement is simply recovered
by specializing such a relation to the ensembles $\mS$ and
$\mR(\mE(\mS))$, with
$\mathrm{f}_1(x):=2K\sqrt{x}\log_2(d_{\s{Q}}/x)$, for sufficiently
small $x$ (typically $x\le2/e^2$).\qquad$\blacksquare$\bigskip


\begin{lemma}
We have that
\begin{equation}
\chi\left(\tilde\mE(\mS)\right)\le\mathrm{f}_2\left(\frac{\sqrt{1-\cfid(\mS,\mE)}}{\zeta(\mS)}\right),
\end{equation}
where $\zeta(\mS)$, for a given ensemble $\mS$, is defined in
Eq.~(\ref{eq:zeta}), and $\mathrm{f}_2(x)$ is a positive, continuous,
monotonic increasing, concave function such that
$\mathrm{f}_2(0)=0$.\qquad$\square$
\end{lemma}
{\bf Proof}. We carefully follow here the proof of Lemma~3
given in Ref.~\cite{rev-extr}. Noticing that we can always take
$d_{\s{A}}\le d_{\s{Q}}^2$, and denoting $\epsilon:=1-\cfid(\mS,\mE)$
and $\epsilon':=\frac{\sqrt{\epsilon}}{\zeta(\mS)} \ge\epsilon$,
provided that $\epsilon'\le 1$ we have that
\begin{equation}\label{eq:tre}
  \chi\left(\tilde\mE(\mS)\right)\le 4N\sqrt{\epsilon'}\log_2\frac{d_{\s{Q}}}{\epsilon'},
\end{equation}
where $N\ge K$ is the number of states in the optimal path in
Eq.~(\ref{eq:eta}), which is always greater of equal than the number
of states $K$ emitted by the source.\qquad$\blacksquare$\bigskip


Finally we obtained the following
\begin{prop}
  Let $\mE:\s{Q}\to\s{Q}'$ be a channel. Let
  $\mS:=\{p(x),|\psi_x\>\}_x$ be an input ensemble of pure
  states. Then, provided that $\cfid(\mS,\mE)\ge 1-(\zeta(\mS))^2$ and
  that $\cfid(\mS,\mE)$ is sufficiently close to one, we have that
\begin{equation}
  \delta(\varrho_\mS,\mE)\le\mathrm{f}\left(\frac{\sqrt{1-\cfid(\mS,\mE)}}{\zeta(\mS)}\right),
\end{equation}
where $\mathrm{f}(x)$ is a positive, continuous, monotonic increasing,
concave function such that $\mathrm{f}(0)=0$. In other words, whenever
the ensemble $\mS$ is irreducible, that means $\zeta(\mS)>0$, the
condition $\cfid(\mS,\mE)\to 1$ implies that
$\delta(\varrho_\mS,\mE)\to 0$.\qquad$\square$
\end{prop}
{\bf Proof}. Simply by combining Eqs.~(\ref{eq:uno}),
(\ref{eq:due}), and (\ref{eq:tre}), and noticing that
$\epsilon'\ge\epsilon$ and $N\ge K$ we see that
\begin{equation}
  \delta(\varrho_\mS,\mE)\le 6N\sqrt{\epsilon'}\log_2\frac{d_{\s{Q}}}{\epsilon'},
\end{equation}
for
\begin{equation}
\epsilon':=\frac{\sqrt{1-\cfid(\mS,\mE)}}{\zeta(\mS)},
\end{equation}
which is the statement.\qquad$\blacksquare$

\end{document}